\documentclass[mnsc,nonblindrev]{informs3_hide} %

\OneAndAHalfSpacedXI %

\usepackage{amsmath,amsfonts,amssymb}
\usepackage{mathtools}
    \allowdisplaybreaks
\usepackage{bm}
\usepackage[mathscr]{euscript}
\usepackage{enumitem}
\usepackage{xcolor}
\usepackage{graphicx}
\usepackage{multirow}
\usepackage{booktabs}
\usepackage{microtype}
\usepackage{fix-cm}
\usepackage[ruled,vlined]{algorithm2e}
\usepackage{physics}
\usepackage{xfrac}
\usepackage{nicematrix}

\usepackage{natbib}
 \bibpunct[, ]{(}{)}{,}{a}{}{,}%
 \def\bibfont{\small}%
\TheoremsNumberedThrough     %
\ECRepeatTheorems

\EquationsNumberedThrough    %

\MANUSCRIPTNO{}

\begin{document}

\RUNAUTHOR{Li et al.}

\RUNTITLE{AgentGit}

\TITLE{AgentGit: A Version Control Framework for Reliable and Scalable LLM-Powered Multi-Agent Systems}

\ARTICLEAUTHORS{
\AUTHOR{Yang Li, Siqi Ping, Xiyu Chen, Xiaojian Qi}
\AFF{University of Hong Kong, Hong Kong SAR}
\AUTHOR{Zigan Wang}
\AFF{Tsinghua University, China}
\AUTHOR{Ye Luo}
\AFF{University of Hong Kong, Pokfulam Road, Hong Kong SAR, \EMAIL{kurtluo@hku.hk}}
\AUTHOR{Xiaowei Zhang}
\AFF{Hong Kong University of Science and Technology, Hong Kong SAR, \EMAIL{xiaoweiz@ust.hk}}
}

\ABSTRACT{With the rapid progress of large language models (LLMs), LLM-powered multi-agent systems (MAS) are drawing increasing interest across academia and industry. However, many current MAS frameworks struggle with reliability and scalability, especially on complex tasks. We present AgentGit, a framework that brings Git-like rollback and branching to MAS workflows. Built as an infrastructure layer on top of LangGraph, AgentGit supports state commit, revert, and branching, allowing agents to traverse, compare, and explore multiple trajectories efficiently.
To evaluate AgentGit, we designed an experiment that optimizes target agents by selecting better prompts. We ran a multi-step A/B test against three baselines---LangGraph, AutoGen, and Agno---on a real-world task: retrieving and analyzing paper abstracts. Results show that AgentGit significantly reduces redundant computation, lowers runtime and token usage, and supports parallel exploration across multiple branches, enhancing both reliability and scalability in MAS development.
This work offers a practical path to more robust MAS design and enables error recovery, safe exploration, iterative debugging, and A/B testing in collaborative AI systems.
}


\maketitle

\section{Introduction}

As large language models (LLMs) advance rapidly \citep{sindhu2024evolution,hadi2023large}, their use in multi-agent systems (MAS) is drawing growing interest from both academia and industry \citep{li2024survey,yang2024llm}. An MAS comprises multiple autonomous agents---often powered by LLMs---that interact within a shared environment to achieve complex goals through coordination and communication. MAS applications span many domains, including automated software development~\citep{qian2023chatdev,wang2024openhands,he2025llm}, scientific simulation~\citep{uhrmacher2018multi}, drug discovery and design~\citep{fu2017designing,kodela2025autonomous}, intelligent transportation~\citep{zhou2022multi,troullinos2021collaborative}, financial analysis~\citep{hafezi2015bat,raudys2006multi}, and intelligent tutoring in education~\citep{vicari2002use}.

However, current LLM-powered MAS fall short of industrial-grade needs, which require high reliability and treat scalability as a core property for future growth~\citep{rana2000scalability,lee1998stability}. Empirical studies report that most proposed MAS achieve low accuracy (often below 50\%) even within their target domains~\citep{pan2025multiagent}, and taxonomies of these failures point to systemic reliability gaps. In addition to reliability, scalability remains a major challenge.

Regarding the poor performance of LLM-powered MAS, we argue that issues like failed tool calls, unexecutable instructions, and endless reasoning loops are symptoms rather than root causes.
The core limitation lies in the architecture: most agent frameworks lack a rollback mechanism.
When execution fails, the system cannot be reverted to a stable state to explore alternative paths.
As a result, a single erroneous action can cascade into full task failure, causing the entire agentic workflow to collapse irreversibly and wasting accumulated context.

Mainstream LLM agent frameworks such as LangChain (\url{https://docs.langchain.com/}) and LangGraph~\citep{pelluru2025langchain} provide modular state control and composable workflows.
However, their execution remains largely linear and irreversible: each agent action mutates state without built-in, lossless recovery.
While LangGraph supports rollback, its mechanism discards intermediate results during restoration, limiting its capability to preserve the full execution context.
When an error occurs, intermediate states are lost.
Without commit/rollback, MAS cannot perform localized recovery, branching exploration, or incremental optimization. As a result, reliability and scalability are constrained in real-world, dynamic environments.

To address this limitation, we propose AgentGit, a framework that introduces Git-like version control semantics into agentic workflows. Built as an infrastructure layer on top of LangGraph, AgentGit provides state commit, state revert, and branching operations that allow agents to traverse, compare, and explore multiple trajectories. With this mechanism, an agent can automatically roll back to its last stable checkpoint upon failure and attempt a different strategy, transforming fragile, linear pipelines into robust, explorable, and self-correcting systems.

In terms of reliability, AgentGit enables fine-grained error recovery, deterministic replay, and safe experimentation. The rollback and branching capabilities allow agents to restore consistent states, isolate faulty trajectories, and re-execute alternative actions without compromising prior results. Furthermore, reproducible checkpoints facilitate systematic debugging, unit testing, and policy validation.

In terms of scalability, AgentGit supports parallel exploration across multiple branches, allowing diverse strategies to evolve independently without redundant computation. Its persistent checkpoint architecture allows large-scale agents to share and reuse prior states across sessions and tasks, thus providing a scalable foundation for building complex, multi-agent ecosystems.

\begin{figure}[ht]
\centering
\includegraphics[width=0.45\textwidth]{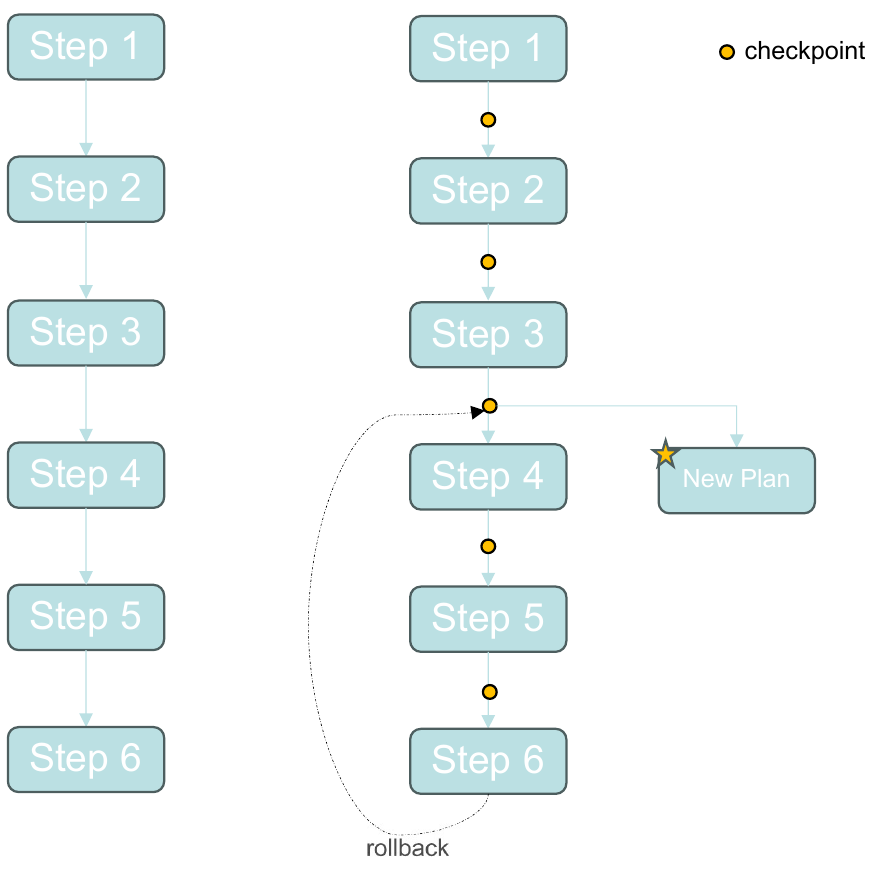}
\caption{Comparison of task execution workflows: standard model vs. AgentGit with rollback functionality}
\label{compare workflow with/without rollback}
\end{figure}

To demonstrate the effectiveness of AgentGit, we designed a set of experiments to compare the efficiency of different frameworks in completing complex tasks. The experiment simulated a real-world scenario: retrieving abstracts of papers from arXiv on a specific topic and performing subsequent analysis and optimization. The workflow consisted of four steps: Search and Extract, Introduction, Analysis, and Discussion. In the experiment, we conducted an A/B test to compare the performance of four frameworks—LangGraph, AutoGen, Agno, and LangGraph+AgentGit—in terms of tool invocation and prompt generation. While all frameworks ultimately arrived at the global optimal solution, the focus of the experiment was to compare their efficiency, particularly in testing different tools and prompts.

The experimental results demonstrated that AgentGit significantly outperformed other frameworks in execution efficiency. By leveraging its unique rollback mechanism, AgentGit allowed skipping previously completed steps and directly testing new tools or prompts from specific nodes, thereby avoiding redundant execution of earlier workflows. Compared to other frameworks, AgentGit substantially reduced the overall runtime and resource consumption, confirming its reliability and scalability in complex task scenarios.

The contributions of this work are as follows:
\begin{itemize}
    \item We propose AgentGit, the first multi-agent framework toolkit that introduces Git-like rollback and branching mechanisms into LLM-powered agent systems, enabling efficient and reversible execution in complex workflows.
    \item We analyze the theoretical complexity of the rollback mechanism in AgentGit, demonstrating its scalability and efficiency in reducing redundant computations during iterative tasks.
    \item We design and conduct an A/B test task to evaluate the performance of AgentGit in comparison to other frameworks (LangGraph, AutoGen, and Agno). Experimental results show that the rollback functionality of AgentGit significantly improves testing efficiency by reducing token consumption and optimizing runtime.
    \item We introduce potential application cases for AgentGit, such as error recovery, safe exploration, iterative debugging, and A/B testing, highlighting its versatility in accelerating MAS development and enhancing system robustness.
    \item We fully open-source our dataset, codebase, and the AgentGit framework to facilitate further research and development in the field of MAS.
\end{itemize}

\section{Related Works}

LLM-powered multi-agent frameworks provide standardized infrastructures for designing, orchestrating, and evaluating interactions among multiple agents powered by LLMs, thus simplify the development of LLM-powered MAS~\citep{li2024survey,he2025llm}. Such frameworks abstract complex coordination processes—such as message passing, role assignment, and tool invocation—into reusable components, enabling researchers and developers to efficiently prototype and deploy collaborative AI systems. In recent years, with the rapid development of LLM technology, the application scenarios of MAS have expanded significantly, including automated software development, scientific simulations, knowledge graph construction, and intelligent customer service systems~\citep{qian2023chatdev,pan2025multiagent}. These scenarios demand higher reliability and scalability from MAS~\citep{rana2000scalability,lee1998stability}.

The current mainstream multi-agent frameworks include LangGraph (\url{https://github.com/langchain-ai/langgraph}), Agno (\url{https://docs.agno.com/introduction}), Autogen~\citep{wu2024autogen}, CrewAI (\url{https://github.com/crewAIInc/crewAI}) and Dify~(\url{https://github.com/langgenius/dify}). LangGraph adopts a graph-based orchestration paradigm, representing agent workflows as directed acyclic graphs. This design ensures deterministic control and reproducibility, making it well-suited for structured pipelines. In contrast, AutoGen models agents as conversational entities that exchange natural language messages, allowing flexible, dialogue-driven coordination among agents and humans. CrewAI organizes agents into role-based teams, where each agent is assigned a specific responsibility under a shared objective—an approach that closely resembles human organizational structures and facilitates multi-role collaboration. Agno focuses on flexible orchestration and adaptive agent coordination, allowing the system to evolve its behavior over time. Dify offers a low-code environment where developers can visually compose and deploy agent workflows, aiming for ease of deployment rather than experimental control.

Although these frameworks exhibit certain advantages in specific scenarios, they share a critical limitation: the lack of rollback mechanisms. Specifically, these frameworks cannot restore to previously stable states during task execution, nor can they support multi-branch exploration or error recovery. Once an agent executes an incorrect action, the overall workflow often fails irreversibly, requiring human intervention. What's more, while LangGraph supports rollback, its mechanism deletes subsequent results upon reverting to a previous state, limiting its ability to retain and reuse intermediate data. This limitation not only increases the cost of task failures but also restricts the applicability of MAS in complex and dynamic environments.

Our work addresses this limitation by introducing a rollback-capable multi-agent framework that enhances robustness through reversible execution and controlled state restoration, enabling more reliable multi-agent collaboration. Furthermore, our framework supports multi-branch exploration and persistent state storage, allowing MAS to perform efficient testing and optimization in complex task scenarios.

\begin{figure}[ht]
\centering
\includegraphics[width=0.35\textwidth]{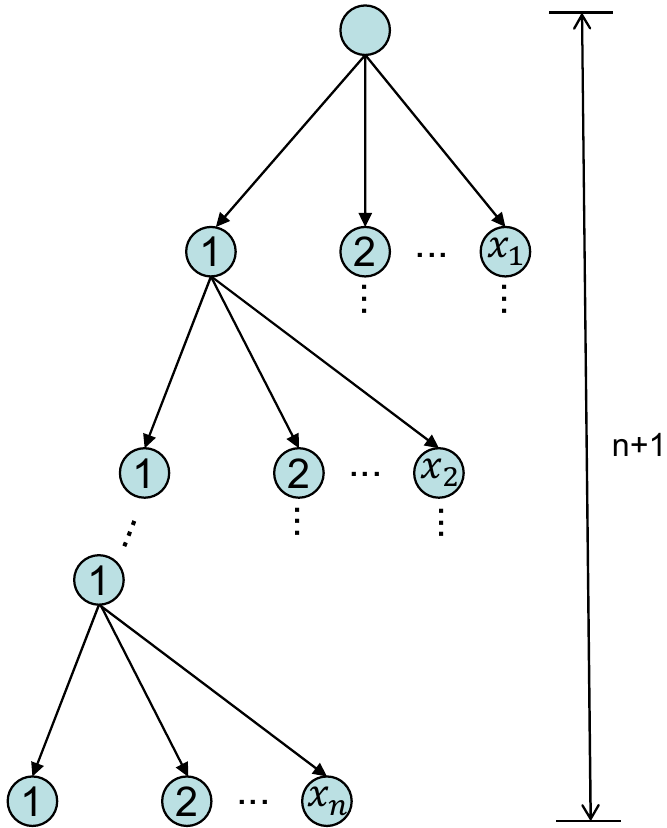}
\caption{Tree diagram illustrating the branching structure of the task execution process}
\label{general tree structure}
\end{figure}

\section{AgentGit}
AgentGit significantly improves the execution efficiency and flexibility of MAS by introducing rollback and branching mechanisms.

\subsection{Rollback}
The rollback mechanism is one of the core features of AgentGit. It creates permanent checkpoints to save the complete system state, including session history, tool invocation records, environment variables, and intermediate reasoning processes. When rollback is required, the system can restore its state from a specified checkpoint, avoiding the need to re-execute previously completed steps. Checkpoints can be created manually or automatically triggered after critical operations. The need for rollback is particularly evident in complex task scenarios, such as testing the effectiveness of different tools or prompts. The rollback mechanism allows the system to directly test new approaches from a specific checkpoint without re-executing earlier steps, significantly reducing overall runtime and resource consumption. The rollback process involves the following steps: first, the system loads the state corresponding to the checkpoint ID specified by the user; next, it restores session history and tool invocation records; finally, it resumes subsequent tasks from the restored state. Through this mechanism, AgentGit effectively optimizes task execution workflows and enhances overall system efficiency.

Figure~\ref{compare workflow with/without rollback} shows the differences between the standard model and AgentGit with rollback mechanism in the task execution process. The standard model in the left executes tasks linearly, with each step permanently altering the system state. If an error occurs or adjustments are needed at Step 4, the system must restart from Step 1, leading to redundant computation and resource waste. AgentGit creates checkpoints after each critical step (e.g., Step 3). If adjustments are needed at Step 4, the system can roll back to the checkpoint at Step 3 and directly test new approaches from that state, without re-executing Steps 1, 2 and 3. This mechanism significantly reduces redundant computation and improves efficiency.

\subsection{Branching}
Branching enables the creation of new branch paths from specific checkpoints. This mechanism allows the system to independently explore different strategies or approaches across multiple paths. Each branch inherits the complete state information of the original path, including session history, tool invocation records, environment variables, and intermediate reasoning processes, ensuring the integrity and independence of each branch. The process of creating a branch involves the following steps: first, the system loads the state from the specified checkpoint; second, a new branch ID is generated, and the branch environment is initialized; finally, users can test different tools or prompt strategies on the new branch without affecting the execution of the original path.

The significant advantage of the branching mechanism is its support for parallel computation. Multiple branches can run simultaneously, testing different strategies or approaches, thereby significantly improving task execution efficiency. For instance, in a complex task, users can test the effectiveness of Tool A on one branch while testing Tool B on another branch, without waiting for one test to complete before starting the next. Additionally, branching supports navigation and merging operations, allowing users to switch between branches, review execution results, and integrate the outcomes of multiple branches. The merging process is similar to Git's branch merging operation, supporting conflict detection and resolution to ensure the completeness and consistency of the final result.

\begin{figure}[ht]
\centering
\includegraphics[width=0.8\textwidth]{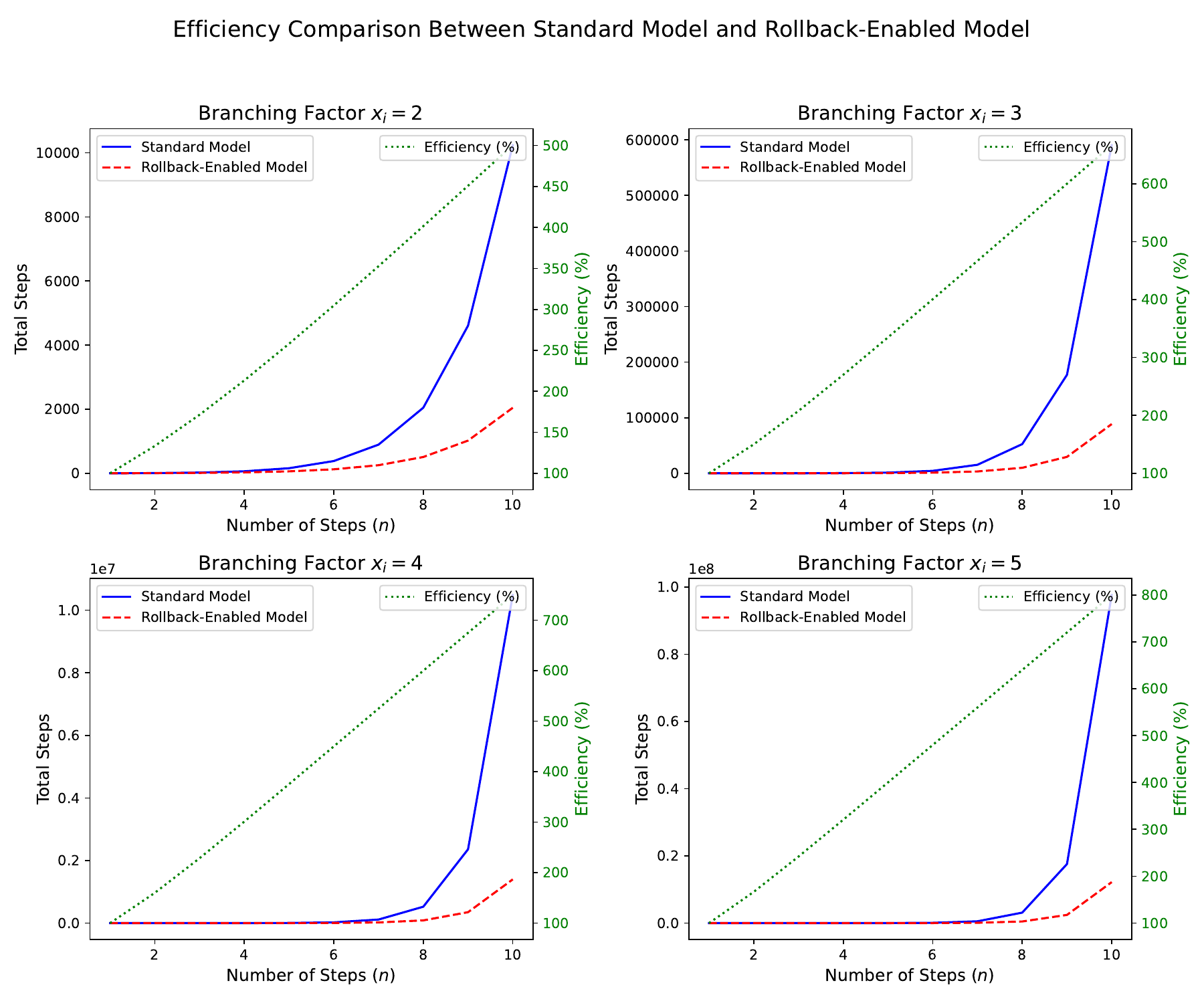}
\caption{Visualization of the total steps required and efficiency trends for the standard model and rollback-enabled model under varying $x_i$ and $n$}
\label{efficiency trend}
\end{figure}

\begin{figure}[ht]
\centering
\includegraphics[width=\textwidth]{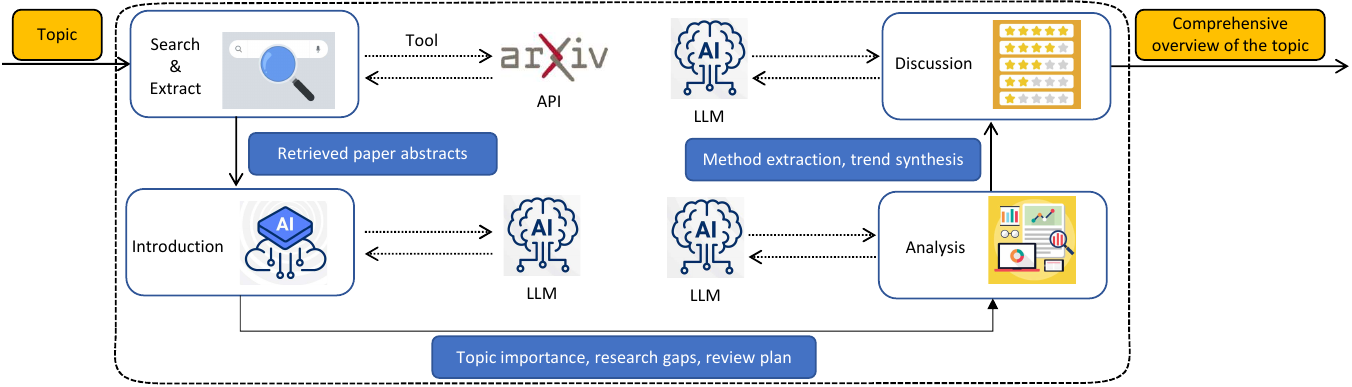}
\caption{Workflow of the MAS task scenario for retrieving abstracts of papers related to a specific topic}
\label{task workflow}
\end{figure}

\subsection{Complexity Analysis}
Complexity analysis demonstrates that AgentGit's mechanism effectively reduces redundant computations and optimizes resource utilization in high-complexity tasks, providing a reliable and efficient solution for the development of MAS.
\begin{lemma}
In an MAS, a workflow consists of \( n \) steps, where each step allows the selection of different tools or prompt options. Suppose the \( i \)-th step has \( x_i \) available tools or prompt options. Then, the total number of possible outcomes \( L \) after executing the workflow can be expressed as:
\[
L = \prod_{i=1}^{n} x_i,
\]
where \( x_i \) represents the number of tools or prompt options available at the \( i \)-th step.
\end{lemma}

To facilitate the understanding of the task execution process and rollback mechanism, we use a tree diagram to represent the branching structure of the entire task, as illustrated in Figure~\ref{general tree structure}. In the tree diagram, the \emph{root node} represents the initial input of the task, such as the original task or question provided by the user. The \emph{intermediate nodes} represent the intermediate results after each step of execution, which can be stored as checkpoints for rollback or branching operations. The \emph{leaf nodes} represent the final results of the task, such as the optimized content or results generated after completing all steps.

The \emph{edges} of the tree represent the specific execution paths from one node to the next, reflecting the choice of tools or LLMs. The \emph{height of the tree} is $n+1$, where $n$ is the number of steps in the task. Each step corresponds to one layer of the tree, and the root node occupies the first layer as the initial input, making the total height of the tree $n+1$. For the tree structure, each node at the $(i-1)$th layer has $x_i$ possible execution paths leading to nodes in the $i$th layer. Therefore, the number of leaf nodes in the tree $L$ can be expressed as:
\[
L = x_1 \cdot x_2 \cdot x_3 \cdot \dots \cdot x_n= \prod_{i=1}^{n} x_i.
\]

\begin{lemma}
In an MAS without rollback mechanism, if the workflow consists of $n$ steps, and the $i$th step has $x_i$ possible tools or prompt options, then the total number of steps required to generate all possible outputs is given by:
\[
\mathcal{S}_{\text{std}} = n \prod_{i=1}^{n} x_i.
\]
\end{lemma}

Let $\mathcal{S}_{\text{std}}$ denote the total steps for the standard model. Since each path contains $n$ steps and the number of leaf nodes in the tree $L= \prod_{i=1}^{n} x_i$. Without a rollback mechanism, the standard model needs to execute the full path for each leaf node, so
\[
\mathcal{S}_{\text{std}} = n  \prod_{i=1}^{n} x_i.
\]

\begin{lemma}
In an MAS with rollback mechanism, if the workflow consists of $n$ steps, and the $i$th step has $x_i$ possible tools or prompt options, then the total number of steps required to generate all possible outputs is given by:
\[
\mathcal{S}_{\text{rollback}} = \sum_{i=1}^{n} \left( \prod_{j=1}^{i-1} x_j \cdot x_i \right).
\]
\end{lemma}

Let $\mathcal{S}_{\text{rollback}}$ denote the total steps for the rollback-enabled model. With a rollback mechanism, the model can avoid re-executing previous steps by rolling back to a checkpoint, thus the total number of steps required to generate all leaf nodes equals the total number of edges in the tree, so
\begin{align*}
   \mathcal{S}_{\text{rollback}} &= x_1 + x_1 \cdot x_2 + x_1 \cdot x_2 \cdot x_3 + \dots + \prod_{i=1}^{n} x_i \nonumber \\
   &= \sum_{i=1}^{n} \left( \prod_{j=1}^{i-1} x_j \cdot x_i \right),
\end{align*}
where $\prod_{j=1}^{i-1} x_j$ represents the number of nodes at the $i-1$th layer, and $x_i$ represents the number of branches for each node.

It indicates significantly from above that the total number of steps required differs largely between a standard MAS and a rollback-enabled one.

Figure~\ref{efficiency trend} visualizes the total steps required for both the standard MAS and the rollback-enabled one. In this figure, we assume that each intermediate node has 2, 3, 4, or 5 child nodes or branches (i.e., $x_i = 2, 3, 4, 5$), and vary the number of steps $n$. From Figure~\ref{efficiency trend}, we can observe that the rollback mechanism significantly reduces the total number of steps required to generate all leaf nodes, especially as the number of steps $n$ increases. This demonstrates the substantial efficiency improvement achieved by AgentGit in complex task scenarios.

\begin{proposition}
    In an MAS with \( n \) steps, if each step has \( \alpha \) possible tools or prompt options (where \( \alpha \) is a constant), define efficiency \( \eta \) as the ratio of the total steps required by a system without rollback mechanism to the total steps required by a system with rollback mechanism to generate all possible final results. When \( N \to \infty \), the growth trend of efficiency \( \eta \) becomes infinite.
\end{proposition}

From above, let $\eta$ denote the efficiency To quantify the efficiency difference between the two mechanisms, defined as the ratio of the total steps required by the standard model to the total steps required by the rollback-enabled model:
\[
   \eta = \frac{\mathcal{S}_{\text{std}}}{\mathcal{S}_{\text{rollback}}}
   =\frac{n  \prod_{i=1}^{n} x_i}{\sum_{i=1}^{n} \left( \prod_{j=1}^{i-1} x_j \cdot x_i \right)}.
\]

Consider a special case where each step has \( \alpha \) possible tools or prompt options. This is equivalent to saying that the number of branches for each intermediate node $x_i$ is a constant $\alpha$. In this scenario, the efficiency $\eta$ can be rewritten as:
\[
\eta = \frac{n \alpha^n}{\sum_{i=1}^{n} \alpha^i}.
\]

When $n$ approaches infinity, we can analyze the growth trend of efficiency using the limit:
\begin{align*}
    \lim_{n \to \infty} \eta &= \lim_{n \to \infty} \frac{n \alpha^n}{\alpha \cdot \frac{\alpha^n - 1}{\alpha - 1}} =\lim_{n \to \infty} n \cdot \frac{\alpha-1}{\alpha}\frac{1}{1 - \alpha^{-n}}\\
    &=\lim_{n \to \infty} n \cdot (1-\alpha^{-1})=\infty.
\end{align*}

This indicates that the growth trend of efficiency $\eta$ is infinite as $n$ increases, and Figure~\ref{efficiency trend} intuitively demonstrates this point.

\begin{proposition}
    In an MAS with \( n \) steps, if each step has \( \alpha \) possible tools or prompt options (where \( \alpha \) is a constant), define efficiency \( \eta \) as the ratio of the total steps required by a system without rollback mechanism to the total steps required by a system with rollback mechanism to generate all possible final results. When \( N \to \infty \), the limit of $\eta / n$ approaches \(\frac{\alpha - 1}{\alpha}\) .
\end{proposition}

$\eta / n$ represents the average efficiency improvement brought by each step of task execution. By calculating $\lim_{n \to \infty} \eta / n$, we can quantify the performance of the rollback mechanism in high-complexity tasks:
\begin{align*}
    \lim_{n \to \infty} \frac{\eta}{n} = &\lim_{n \to \infty} \frac{\alpha^n}{\sum_{i=1}^{n} \alpha^i} = \lim_{n \to \infty} \frac{\alpha^n}{\alpha \cdot \frac{\alpha^n - 1}{\alpha - 1}}\\
    &=\lim_{n \to \infty} \frac{\alpha-1}{\alpha}\frac{1}{1 - \alpha^{-n}}= \frac{\alpha - 1}{\alpha}.
\end{align*}

As the task complexity increases, the average efficiency improvement per unit task complexity stabilizes, with its limit given by:
\[
\lim_{n \to \infty} \frac{\eta}{n} = \frac{\alpha - 1}{\alpha}.
\]

This result indicates that the impact of the number of branches $\alpha$ on efficiency improvement diminishes as $\alpha$ increases, and the average efficiency improvement approaches 1 when $\alpha$ tends to infinity. Figure~\ref{efficiency trend} visually illustrates this observation. From the curves in the figure, it can be seen that their slopes gradually stabilize as \( N \to \infty \).

\begin{figure}[ht]
\centering
\includegraphics[width=0.5\textwidth]{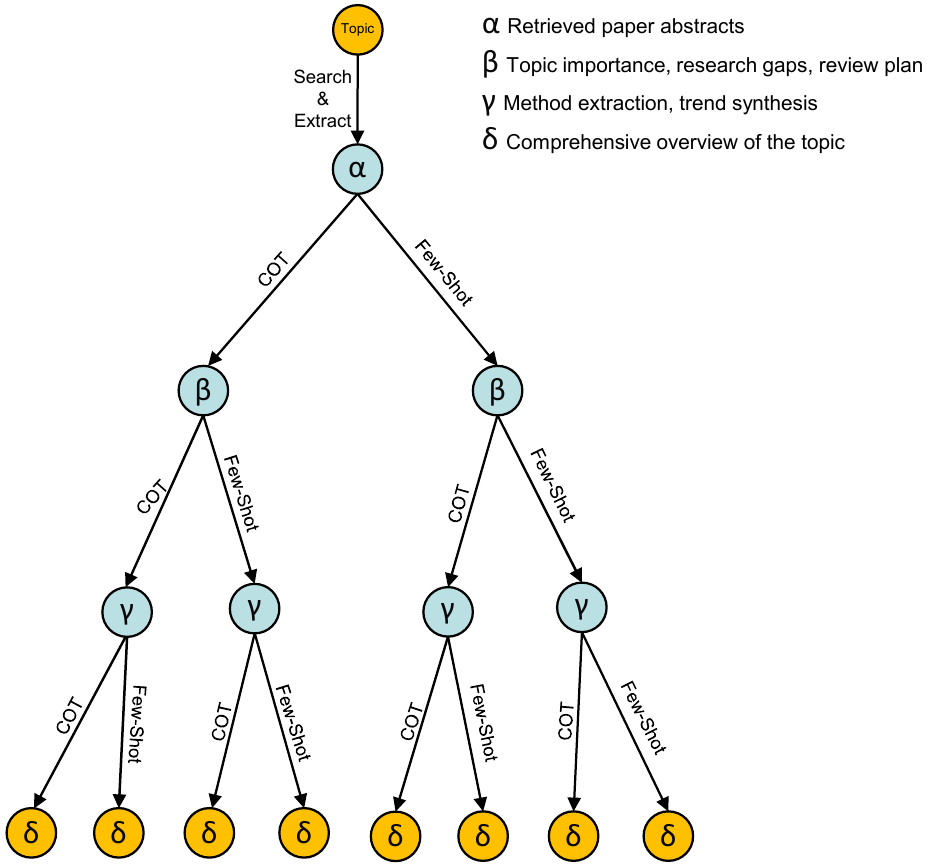}
\caption{Tree structure representing the experimental workflow}
\label{experiment tree structure}
\end{figure}

\section{Experiment}
While AgentGit has a wide range of potential applications, such as error recovery, safe exploration, iterative debugging, and A/B testing, we chose a representative experiment to demonstrate its efficiency in complex task scenarios—retrieving abstracts of papers from arXiv on a specific topic and performing subsequent analysis and optimization. In this experiment, we conducted an A/B test to compare the effectiveness of different prompt generation methods, aiming to identify the optimal combination for completing the task.

\begin{figure}[ht]
\centering
\includegraphics[width=0.6\textwidth]{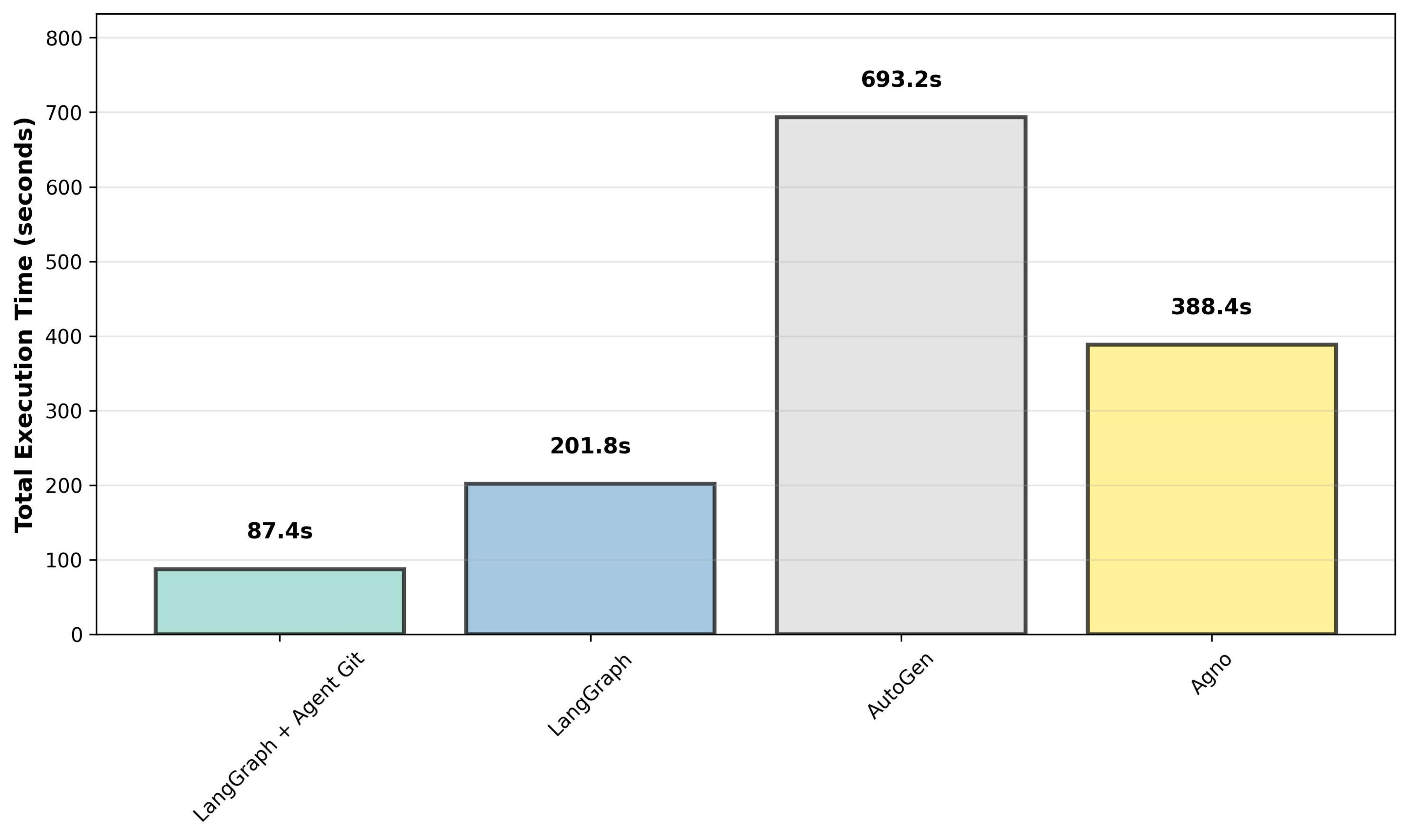}
\caption{Execution time comparison of different frameworks for completing the task}
\label{Execution Time}
\end{figure}

\subsection{Experimental Setup}
\subsubsection{Task Scenario} This experiment simulates an MAS task scenario, where the task is to retrieve abstracts of papers related to a specific topic from arXiv, analyze and optimize these abstracts, and generate a final comprehensive report. The workflow of the task scenario is illustrated in Figure \ref{task workflow}. This task aims to identify the optimal prompt generation methods through A/B test and consists of four steps: Search and Extract, Introduction, Analysis, and Discussion. First, the system retrieves paper titles and abstracts related to the specified topic by invoking arXiv API. Then, it filters and extracts abstracts from the retrieved papers to ensure relevance and uniqueness. Next, the system generates an Introduction based on the extracted abstracts using different prompt generation methods. Following this, the system performs an Analysis of the abstracts, extracting key insights and evaluating their contributions. Finally, the system generates a Discussion section to summarize the findings and provide a comprehensive overview of the topic.
\subsubsection{Baselines} The experiment compares four frameworks: LangGraph, AutoGen, LangGraph+AgentGit, and Agno. These frameworks represent different approaches to MAS design, with LangGraph and AutoGen serving as baseline frameworks, Agno focusing on task decomposition and collaboration, and LangGraph+AgentGit introducing rollback mechanisms to optimize task execution. These frameworks were selected to provide a comprehensive comparison of tool invocation and prompt generation efficiency.

\begin{figure}[b]
\centering
\includegraphics[width=0.6\textwidth]{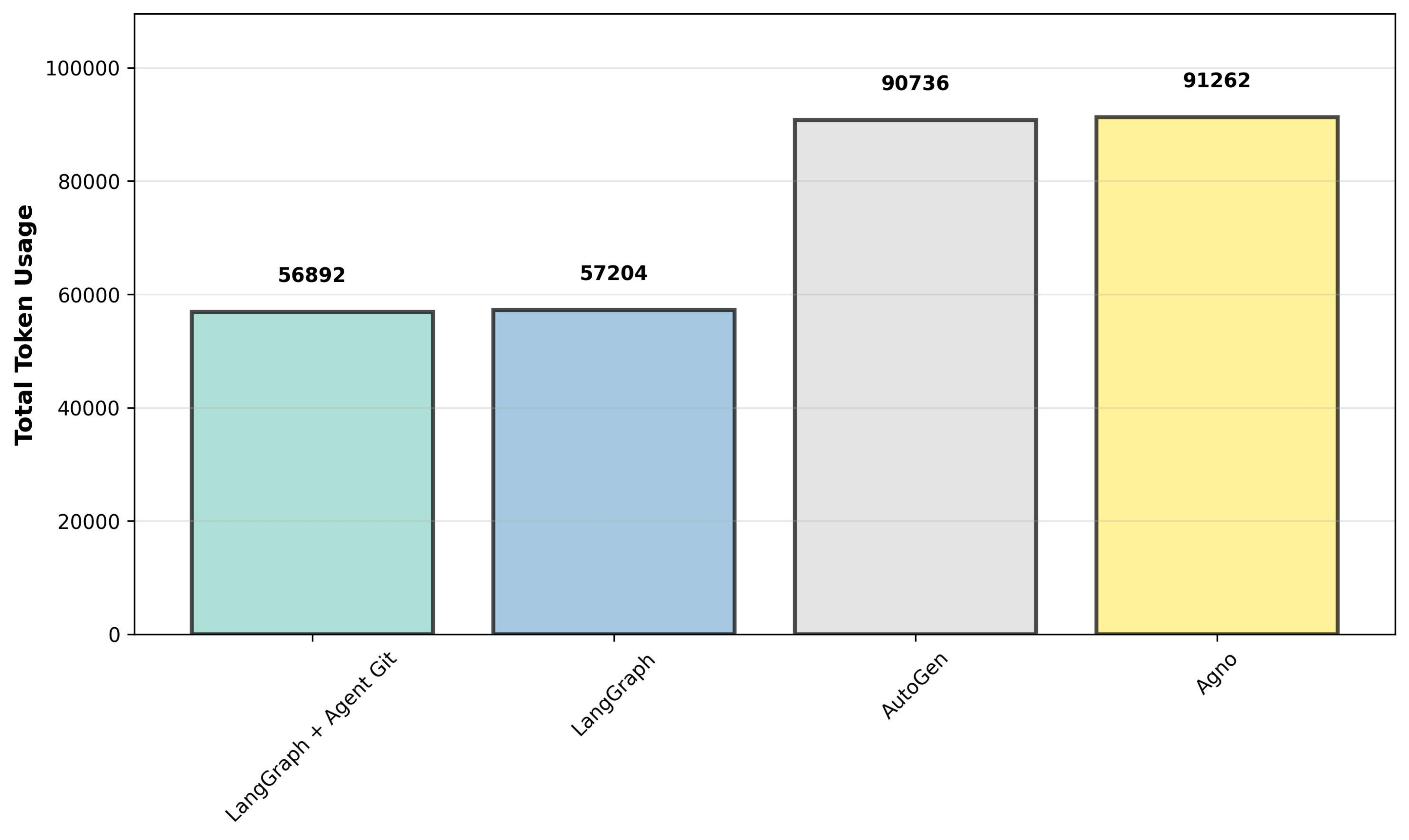}
\caption{Token usage comparison of different frameworks for completing the task}
\label{Token Usage}
\end{figure}

\subsubsection{Experimental Design} The experiment conducts A/B test to compare the effectiveness of different prompt generation methods to identify the optimal approach for completing the task. Specifically, in the Search and Extract step, the system utilized the arXiv API to retrieve paper abstracts related to the specified topic. In the subsequent steps—Introduction, Analysis, and Discussion—prompts generated by different methods, including COT Prompt~\citep{wei2022chain,wei2022emergent} and Few-Shot Prompt~\citep{mann2020language}, were compared to evaluate their impact on task performance.

For this experiment, we used LangGraph version 0.6.6 (released on August 20, 2025) as the baseline framework, and AgentGit version 0.0.1 was integrated into LangGraph to enable rollback mechanisms. Additionally, all frameworks utilized the GPT-4o-mini model with a temperature setting of 0 to ensure consistent outputs for identical inputs.

To evaluate the quality of the final outputs generated by each framework, we employed G-Eval~\citep{liu2023g}, a widely used evaluation metric for assessing the coherence, relevance, and overall quality of generated text. G-Eval scores were assigned to the comprehensive reports produced by each framework under specific prompt combinations, such as COT-COT-COT and Few-Shot-Few-Shot-Few-Shot. This scoring method ensured an objective comparison of the frameworks' ability to generate high-quality outputs.

The experimental design is illustrated in Figure~\ref{experiment tree structure}, which represents the workflow as a tree structure. The root node corresponds to the initial input provided to the MAS, such as the specified topic for retrieving paper abstracts. Each edge in the tree represents a specific operation, such as "Search and Extract with arXiv API" or "Analysis with prompt generated by COT method." Intermediate nodes represent checkpoints, where the system stores the state of the task after completing a specific step. The leaf nodes correspond to the final polished outputs, which are comprehensive reports containing the refined abstracts of papers. This tree structure visually captures the branching possibilities at each step, allowing for systematic exploration of different tools and prompt generation methods.

\begin{figure}[ht]
\centering
\includegraphics[width=\textwidth]{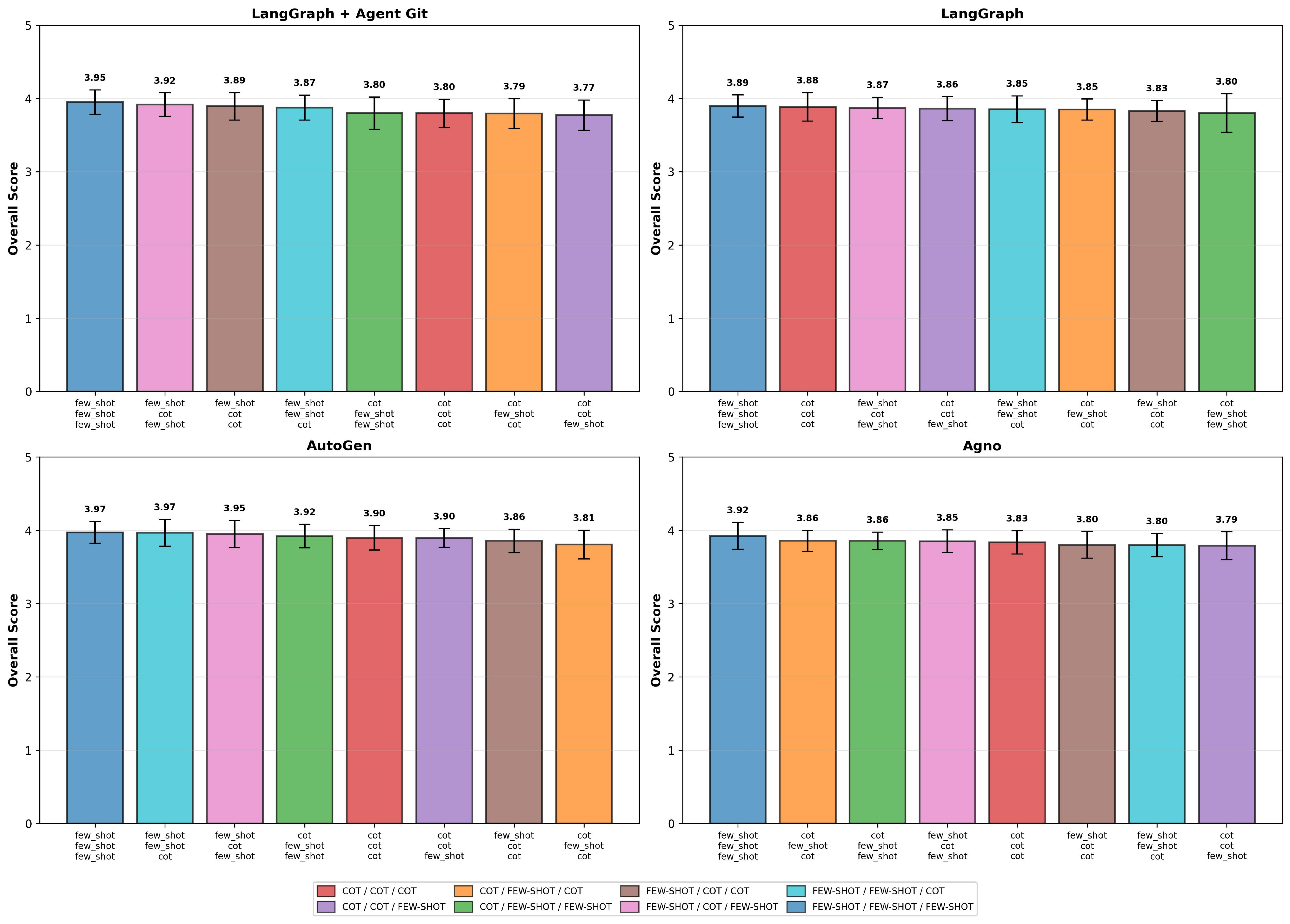}
\caption{G-Eval scores of final outputs generated by specific prompt combinations in different frameworks}
\label{G-Eval scores}
\end{figure}

\subsection{Results}
\subsubsection{Execution Time Analysis}
Figure \ref{Execution Time} illustrates the execution time of the four frameworks for completing the task. LangGraph+AgentGit significantly outperformed the other frameworks in terms of runtime, achieving the shortest execution time. This improvement can be attributed to the rollback and branching mechanism introduced by AgentGit, which allows the system to avoid re-executing previously completed steps and process in parallel. In contrast, LangGraph supports rollback but lacks branching capabilities, requiring sequential execution for each branch tested. AutoGen and Agno, on the other hand, lack both rollback and branching mechanisms, resulting in repeated execution of all four steps for every new test. This result highlights the efficiency of AgentGit in optimizing task execution in complex scenarios.

\subsubsection{Token Consumption Analysis}
Figure \ref{Token Usage} compares the token consumption of the four frameworks during task execution. The experimental results show that LangGraph+AgentGit consumed significantly fewer tokens than AutoGen and Agno, and slightly fewer than LangGraph, remaining almost consistent with it. This result aligns with intuition, as LangGraph also utilizes its rollback mechanism (although it does not permanently store intermediate processes), thereby avoiding the repeated generation of large amounts of content. However, the token consumption of LangGraph+AgentGit is not entirely identical to LangGraph, which may be due to the variance in output generation by large models. Specifically, for the same input, the output may exhibit slight variations, leading to differences in token usage.

\subsubsection{Prompt Combination Performance}
Figure \ref{G-Eval scores} presents the performance scores of different frameworks under specific prompt combinations, such as COT-COT-COT. The experimental results show that for specific prompt combinations, the scores of the four frameworks vary slightly but remain generally consistent. Additionally, for different prompt combinations, the scores within the same framework are also largely consistent. This may indicate that the effects of COT Prompt and Few-Shot Prompt do not differ significantly. The slight variations in scores across frameworks may be attributed to the inherent variance in large language models, which can introduce subtle differences in output generation for identical inputs. While further investigation is needed to fully confirm the consistency of the experimental results, it can be observed that the performance of all four frameworks under different prompt combinations is generally consistent. This consistency further validates the robustness of the frameworks under different prompt combinations, while also demonstrating that LangGraph+AgentGit can maintain output quality comparable to other frameworks while improving execution efficiency.

\section{Conclusion}
In this paper, we introduced AgentGit, a novel framework that integrates Git-like rollback and branching mechanisms into LLM-powered MAS. By enabling state commit, state revert, and branching operations, AgentGit addresses critical limitations in reliability and scalability faced by existing MAS frameworks. Our experimental results demonstrate that AgentGit significantly improves execution efficiency by reducing redundant computations, optimizing runtime, and minimizing token consumption. These findings validate the effectiveness of AgentGit in complex task scenarios, such as retrieving and analyzing paper abstracts.

Beyond the scope of the experiments presented, AgentGit opens up new possibilities for MAS development and application. Its rollback and branching capabilities provide robust solutions for error recovery, safe exploration, iterative debugging, and A/B testing, enabling developers to systematically optimize workflows and enhance system reliability. By supporting parallel exploration and persistent state storage, AgentGit lays the groundwork for scalable and adaptive multi-agent ecosystems, paving the way for future advancements in collaborative AI systems.

\bibliographystyle{informs2014}
\bibliography{ai-conf}

\begin{thebibliography}{24}
\providecommand{\natexlab}[1]{#1}
\providecommand{\url}[1]{\texttt{#1}}
\providecommand{\urlprefix}{URL }

\bibitem[{Brown et~al.(2020)Brown, Mann, Ryder, Subbiah, Kaplan, Dhariwal, Neelakantan, Shyam, Sastry, Askell, Agarwal, Herbert-Voss, Krueger, Henighan, Child, Ramesh, Ziegler, Wu, Winter, Hesse, Chen, Sigler, Litwin, Gray, Chess, Clark, Berner, McCandlish, Radford, Sutskever, \protect\BIBand{} Amodei}]{mann2020language}
Brown T, Mann B, Ryder N, Subbiah M, Kaplan JD, Dhariwal P, Neelakantan A, Shyam P, Sastry G, Askell A, Agarwal S, Herbert-Voss A, Krueger G, Henighan T, Child R, Ramesh A, Ziegler D, Wu J, Winter C, Hesse C, Chen M, Sigler E, Litwin M, Gray S, Chess B, Clark J, Berner C, McCandlish S, Radford A, Sutskever I, Amodei D (2020) Language models are few-shot learners. \emph{Advances in Neural Information Processing Systems 33}, 1877--1901.

\bibitem[{Fu et~al.(2017)Fu, Sun, Sheng, \protect\BIBand{} Liao}]{fu2017designing}
Fu Rg, Sun Y, Sheng Wb, Liao Df (2017) Designing multi-targeted agents: An emerging anticancer drug discovery paradigm. \emph{European Journal of Medicinal Chemistry} 136:195--211.

\bibitem[{Hadi et~al.(2023)Hadi, Qureshi, Shah, Irfan, Zafar, Shaikh, Akhtar, Wu, Mirjalili et~al.}]{hadi2023large}
Hadi MU, Qureshi R, Shah A, Irfan M, Zafar A, Shaikh MB, Akhtar N, Wu J, Mirjalili S, et~al. (2023) Large language models: A comprehensive survey of its applications, challenges, limitations, and future prospects. \url{https://www.techrxiv.org/doi/full/10.36227/techrxiv.23589741.v3}.

\bibitem[{Hafezi et~al.(2015)Hafezi, Shahrabi, \protect\BIBand{} Hadavandi}]{hafezi2015bat}
Hafezi R, Shahrabi J, Hadavandi E (2015) A bat-neural network multi-agent system (bnnmas) for stock price prediction: Case study of dax stock price. \emph{Applied Soft Computing} 29:196--210.

\bibitem[{He et~al.(2025)He, Treude, \protect\BIBand{} Lo}]{he2025llm}
He J, Treude C, Lo D (2025) Llm-based multi-agent systems for software engineering: Literature review, vision, and the road ahead. \emph{ACM Transactions on Software Engineering and Methodology} 34(5):1--30.

\bibitem[{Kodela(2025)}]{kodela2025autonomous}
Kodela KC (2025) Autonomous agentic {AI} systems for pharmaceutical drug discovery: A multi-agent framework for molecular design and optimization. \url{https://ssrn.com/abstract=5382801}.

\bibitem[{Lee et~al.(1998)Lee, Nwana, Ndumu, \protect\BIBand{} De~Wilde}]{lee1998stability}
Lee LC, Nwana HS, Ndumu DT, De~Wilde P (1998) The stability, scalability and performance of multi-agent systems. \emph{BT Technology Journal} 16(3):94--103.

\bibitem[{Li et~al.(2024)Li, Wang, Zeng, Wu, \protect\BIBand{} Yang}]{li2024survey}
Li X, Wang S, Zeng S, Wu Y, Yang Y (2024) A survey on {LLM}-based multi-agent systems: Workflow, infrastructure, and challenges. \emph{Vicinagearth} 1(1):9.

\bibitem[{Liu et~al.(2023)Liu, Iter, Xu, Wang, Xu, \protect\BIBand{} Zhu}]{liu2023g}
Liu Y, Iter D, Xu Y, Wang S, Xu R, Zhu C (2023) {G-Eval}: {NLG} evaluation using {GPT-4} with better human alignment. \url{https://arxiv.org/abs/2303.16634}.

\bibitem[{Pan et~al.(2025)Pan, Cemri, Agrawal, Yang, Chopra, Tiwari, Keutzer, Parameswaran, Ramchandran, Klein et~al.}]{pan2025multiagent}
Pan MZ, Cemri M, Agrawal LA, Yang S, Chopra B, Tiwari R, Keutzer K, Parameswaran A, Ramchandran K, Klein D, et~al. (2025) Why do multiagent systems fail? \emph{ICLR 2025 Workshop on Building Trust in Language Models and Applications}.

\bibitem[{Pelluru(2025)}]{pelluru2025langchain}
Pelluru K (2025) {LangChain} \& {LangGraph} in production: Architectures for multi-agent {LLM} systems. \emph{Journal of Data and Digital Innovation} 2(3):1--9.

\bibitem[{Qian et~al.(2023)Qian, Liu, Liu, Chen, Dang, Li, Yang, Chen, Su, Cong, Xu, Li, Liu, \protect\BIBand{} Sun}]{qian2023chatdev}
Qian C, Liu W, Liu H, Chen N, Dang Y, Li J, Yang C, Chen W, Su Y, Cong X, Xu J, Li D, Liu Z, Sun M (2023) {ChatDev}: Communicative agents for software development. \url{https://arxiv.org/abs/2307.07924}.

\bibitem[{Rana \protect\BIBand{} Stout(2000)}]{rana2000scalability}
Rana OF, Stout K (2000) What is scalability in multi-agent systems? \emph{Proceedings of the Fourth International Conference on Autonomous Agents}, 56--63.

\bibitem[{Raudys \protect\BIBand{} Zliobaite(2006)}]{raudys2006multi}
Raudys {\v{S}}, Zliobaite I (2006) The multi-agent system for prediction of financial time series. \emph{International Conference on Artificial Intelligence and Soft Computing}, 653--662 (Springer).

\bibitem[{Sindhu et~al.(2024)Sindhu, Prathamesh, Sameera, \protect\BIBand{} KumaraSwamy}]{sindhu2024evolution}
Sindhu B, Prathamesh R, Sameera M, KumaraSwamy S (2024) The evolution of large language model: Models, applications and challenges. \emph{2024 International Conference on Current Trends in Advanced Computing (ICCTAC)}, 1--8 (IEEE).

\bibitem[{Troullinos et~al.(2021)Troullinos, Chalkiadakis, Papamichail, \protect\BIBand{} Papageorgiou}]{troullinos2021collaborative}
Troullinos D, Chalkiadakis G, Papamichail I, Papageorgiou M (2021) Collaborative multiagent decision making for lane-free autonomous driving. \emph{Proceedings of the 20th International Conference on Autonomous Agents and Multiagent Systems}, 1335--1343.

\bibitem[{Uhrmacher \protect\BIBand{} Weyns(2018)}]{uhrmacher2018multi}
Uhrmacher AM, Weyns D (2018) \emph{Multi-Agent systems: Simulation and Applications} (CRC press).

\bibitem[{Vicari \protect\BIBand{} Giraffa(2002)}]{vicari2002use}
Vicari RM, Giraffa LMM (2002) The use of multi‐agent systems to build intelligent tutoring systems. \emph{AIP Conference Proceedings} 627(1):340--348.

\bibitem[{Wang et~al.(2025)Wang, Li, Song, Xu, Tang, Zhuge, Pan, Song, Li, Singh, Tran, Li, Ma, Zheng, Qian, Shao, Muennighoff, Zhang, Hui, Lin, Brennan, Peng, Ji, \protect\BIBand{} Neubig}]{wang2024openhands}
Wang X, Li B, Song Y, Xu FF, Tang X, Zhuge M, Pan J, Song Y, Li B, Singh J, Tran HH, Li F, Ma R, Zheng M, Qian B, Shao Y, Muennighoff N, Zhang Y, Hui B, Lin J, Brennan R, Peng H, Ji H, Neubig G (2025) {OpenHands}: An open platform for {AI} software developers as generalist agents. \emph{The Thirteenth International Conference on Learning Representations}, \urlprefix\url{https://openreview.net/forum?id=OJd3ayDDoF}.

\bibitem[{Wei et~al.(2022{\natexlab{a}})Wei, Tay, Bommasani, Raffel, Zoph, Borgeaud, Yogatama, Bosma, Zhou, Metzler, Chi, Hashimoto, Vinyals, Liang, Dean, \protect\BIBand{} Fedus}]{wei2022emergent}
Wei J, Tay Y, Bommasani R, Raffel C, Zoph B, Borgeaud S, Yogatama D, Bosma M, Zhou D, Metzler D, Chi EH, Hashimoto T, Vinyals O, Liang P, Dean J, Fedus W (2022{\natexlab{a}}) Emergent abilities of large language models. \url{https://arxiv.org/abs/2206.07682}.

\bibitem[{Wei et~al.(2022{\natexlab{b}})Wei, Wang, Schuurmans, Bosma, Xia, Chi, Le, \protect\BIBand{} Zhou}]{wei2022chain}
Wei J, Wang X, Schuurmans D, Bosma M, Xia F, Chi E, Le QV, Zhou D (2022{\natexlab{b}}) Chain-of-thought prompting elicits reasoning in large language models. \emph{Advances in Neural Information Processing Systems 35}, 24824--24837.

\bibitem[{Wu et~al.(2024)Wu, Bansal, Zhang, Wu, Li, Zhu, Jiang, Zhang, Zhang, Liu, Awadallah, White, Burger, \protect\BIBand{} Wang}]{wu2024autogen}
Wu Q, Bansal G, Zhang J, Wu Y, Li B, Zhu E, Jiang L, Zhang X, Zhang S, Liu J, Awadallah AH, White RW, Burger D, Wang C (2024) {AutoGen}: Enabling next-gen {LLM} applications via multi-agent conversations. \emph{First Conference on Language Modeling}, \urlprefix\url{https://openreview.net/forum?id=BAakY1hNKS}.

\bibitem[{Yang et~al.(2024)Yang, Peng, Wang, Wen, \protect\BIBand{} Zhang}]{yang2024llm}
Yang Y, Peng Q, Wang J, Wen Y, Zhang W (2024) {LLM}-based multi-agent systems: Techniques and business perspectives. \url{https://arxiv.org/abs/2411.14033}.

\bibitem[{Zhou et~al.(2022)Zhou, Chen, Yan, Li, Yin, \protect\BIBand{} Ge}]{zhou2022multi}
Zhou W, Chen D, Yan J, Li Z, Yin H, Ge W (2022) Multi-agent reinforcement learning for cooperative lane changing of connected and autonomous vehicles in mixed traffic. \emph{Autonomous Intelligent Systems} 2(1):5.

\end{thebibliography}

\end{document}